\documentclass[conference]{IEEEtran}

\usepackage{multirow}
\usepackage{graphicx}
\usepackage[table,xcdraw]{xcolor}
\usepackage{xspace}
\usepackage[english]{babel}
\usepackage{titlesec}
\usepackage{multicol}
\makeatletter
\def\footnoterule{\relax%
  \kern-5pt
  \hbox to \columnwidth{\hfill\vrule width 0.8\columnwidth height 0.4pt\hfill}
  \kern4.6pt}
\makeatother
\usepackage{booktabs}
\usepackage{amsmath}
\usepackage{algorithm}
\usepackage{algpseudocode}
\usepackage{amssymb}
\usepackage{enumitem}
\usepackage{caption}
\usepackage{listings}
\usepackage{subcaption}  
\usepackage{xcolor}
\usepackage{tcolorbox}
\tcbuselibrary{listings} 
\usepackage{hyperref}
\usepackage{cite}

\definecolor{keyword}{RGB}{0,0,255} 
\definecolor{identifier}{RGB}{0,0,0} 
\definecolor{comment}{RGB}{0,128,0} 
\definecolor{string}{RGB}{163,21,21} 
\definecolor{highlight2}{RGB}{255,255,0} 

\renewcommand\thesection{\Roman{section}} 
\renewcommand\thesubsection{\Alph{subsection}}
\renewcommand\thesubsubsection{\thesubsection.\arabic{subsubsection}}

\newcommand{\BibTeX}{{\rm B\kern-.05em{\sc i\kern-.025em b}\kern-.08em
    T\kern-.1667em{\lower.7ex\hbox{E}}\kern-.125emX}}

\begin{document}

\title{Runtime Detection of Adversarial Attacks in AI Accelerators Using Performance Counters}

\author{
    \IEEEauthorblockN{Habibur Rahaman, Atri Chatterjee, and Swarup Bhunia}  
    \IEEEauthorblockA{School of Electrical and Computer Engineering, University of Florida \\  
    \{rahaman.habibur, a.chatterjee\}@ufl.edu, swarup@ece.ufl.edu}  
}

\maketitle

\maketitle

\begin{abstract}

Rapid adoption of AI technologies raises several major security concerns, including the risks of adversarial perturbations, which threaten the confidentiality and integrity of AI applications. Protecting AI hardware from misuse and diverse security threats is a challenging task. To address this challenge, we propose SAMURAI, a novel framework for safeguarding against malicious usage of AI hardware and its resilience to attacks. SAMURAI introduces an AI Performance Counter (APC) for tracking dynamic behavior of an AI model coupled with an on-chip Machine Learning (ML) analysis engine, known as TANTO (\textit{Trained Anomaly Inspection Through Trace Observation}).  APC records the runtime profile of the low-level hardware events of different AI operations. Subsequently, the summary information recorded by the APC is processed by TANTO to efficiently identify potential security breaches and ensure secure, responsible use of AI. SAMURAI enables real-time detection of security threats and misuse without relying on traditional software-based solutions that require model integration.

Experimental results demonstrate that SAMURAI achieves up to 97\% accuracy in detecting adversarial attacks with moderate overhead on various AI models, significantly outperforming conventional software-based approaches. It enhances security and regulatory compliance, providing a comprehensive solution for safeguarding AI against emergent threats.

\end{abstract}


\renewcommand\IEEEkeywordsname{Keywords}


\section{\textbf{Introduction}}



Artificial intelligence (AI) hardware cores, known as AI accelerators, often integrated into modern System-on-Chip (SoC) architecture, provide enhanced performance for AI and machine learning tasks, and facilitate the rapid development and deployment of AI solutions. However, the widespread application of AI also presents several major attack vectors that target its inherent vulnerabilities, since AI often handles sensitive and critical information \cite{kong2021survey,osoba2017risks,yigitcanlar2020contributions}. To ensure security, AI hardware manufacturers must regulate chip usage in terms of duration, operations, and user access, preventing misuse that could lead to cyberattacks, defects, or unauthorized deployment in the cloud \cite{kaloudi2020ai}. Monitoring AI chip behavior is crucial for maintaining secure and safe operations and protecting against potential security breaches and safety hazards.


Generally, the performance and security analysis of AI chips is primarily performed at the software level, which cannot proactively prevent misuse and security threats. They cannot enforce restrictions at the hardware level, which is crucial for preventing unauthorized access/misuse and addressing the security issue of AI cores. Again, the security and usage analysis of AI chips is monitored by recording the occurrence of different metrics of the processor \cite{kumar2021inferring} by the hardware performance counter (HPC) \cite{das2019sok} built in with the processor. HPCs consisting of several numbers of counters count these events occurring within the processor and monitor general system performance metrics of threat detection in AI.

Various malicious attacks \cite{madry2017towards, moosavi2016deepfool,mittal2021survey, hong2018security, rakin2019bit, fredrikson2015model,liang2022adversarial} on neural networks (NN) target vulnerabilities in models. 
Among the main attack models, the percentage occurrence of adversarial attacks is around 36\%, which is the highest, causing increasing concern in the models. Although DNNs show strong performance in classification tasks, they are highly vulnerable to adversarial perturbations of the data \cite{moosavi2016deepfool}. Even small and imperceptible changes to the input data can cause the models to misclassify, exposing a critical weakness in their stability.

\begin{figure}[!h]
    \centering
    \includegraphics[width=0.45\textwidth]{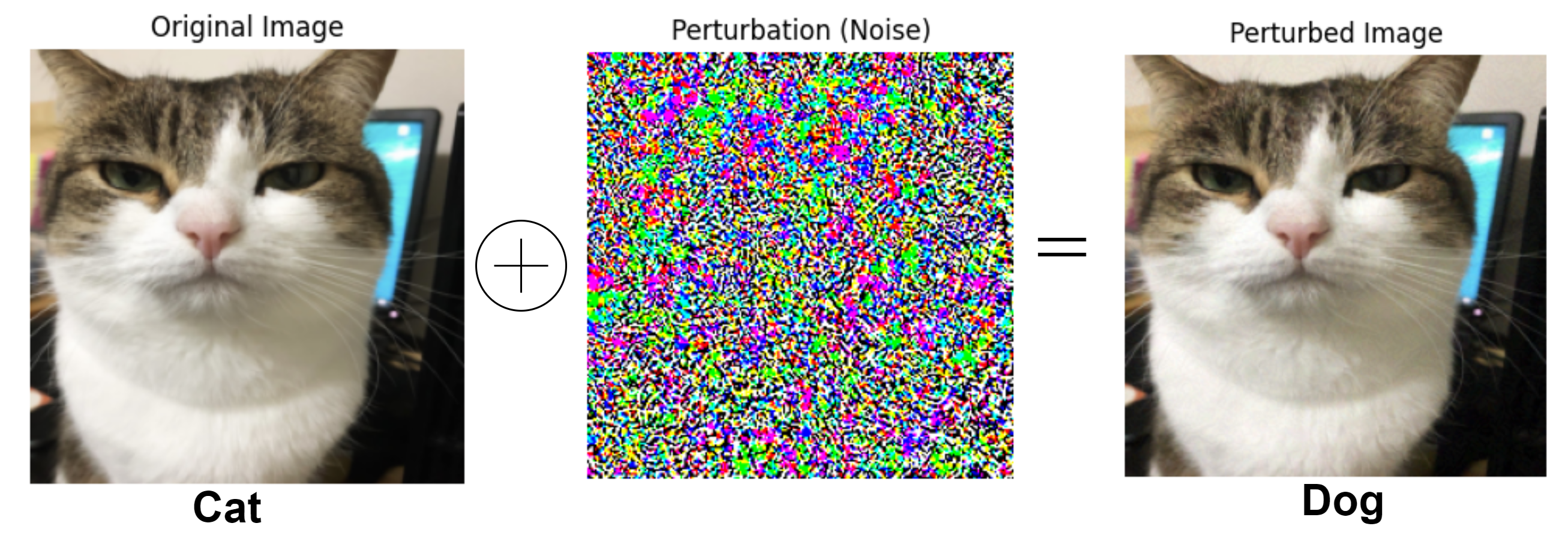}
    \caption {An example of deep-fool attack on an ML model.}
    \label{fig:A1}
\end{figure}

An example of DeepFool adversarial perturbations is illustrated in Figure \ref{fig:A1}. The original image is classified as ``Cat" and the third image incorrectly identifies ``Cat" as ``Dog" due to addition of small perturbation (second image) caused by the DeepFool attack, although difference between original image and adversarial image cannot be detected by human eyes.

\begin{figure}[!h]
    \centering
    \includegraphics[width=0.45\textwidth]{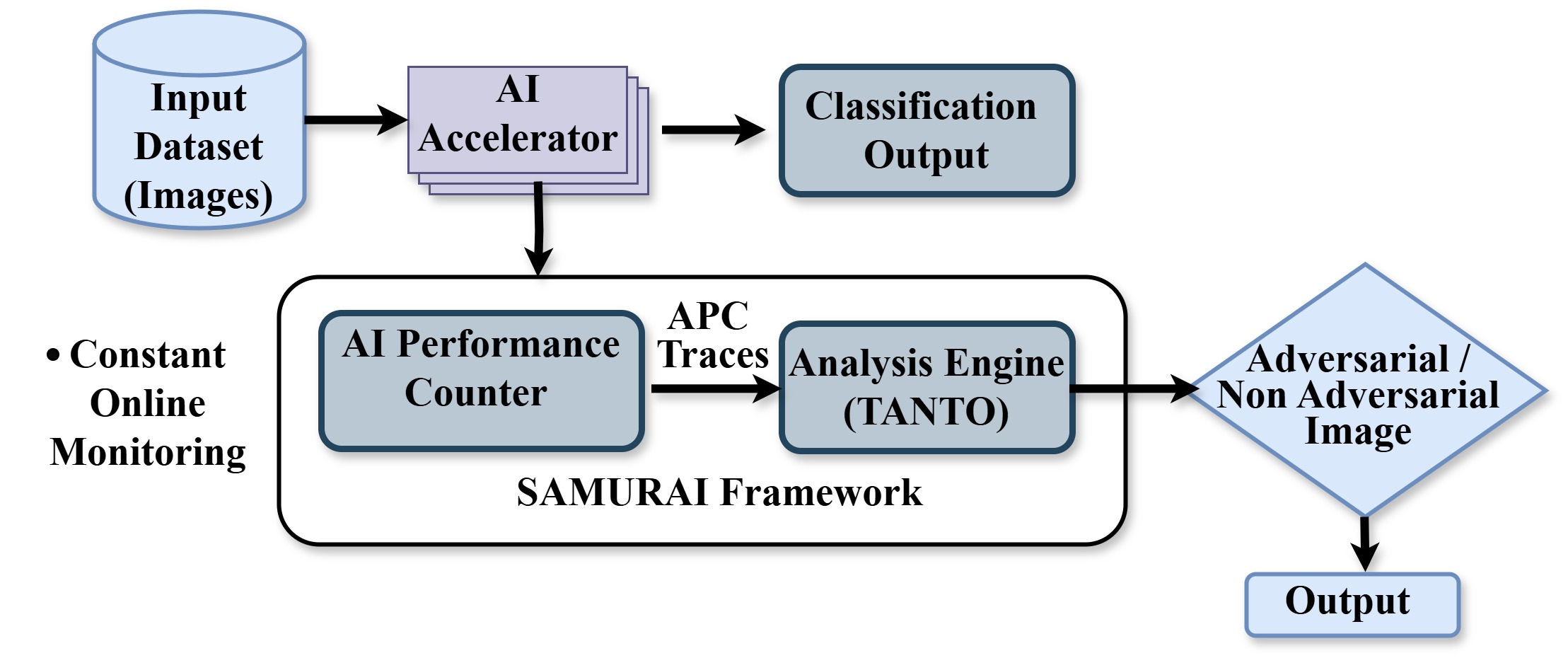}
     \caption{Block Diagram of the SAMURAI Framework illustrating the integration of the AI Performance Counter (APC) and the TANTO analysis engine, where APC captures AI hardware activity and TANTO processes the data in real time to detect adversarial attacks and unauthorized usage.}
    \label{fig:A2}
\end{figure}


The existing techniques do not effectively monitor AI operations and detect AI-specific threats, as these do not capture AI-specific metrics like tensor operations, layer details, or specific AI parameters. As HPCs typically monitor basic system-level metrics, they may not reflect the anomalies specific to AI operations. These methods are not effective for detecting different attacks, mainly adversarial attacks. The DeepFool attack \cite{moosavi2016deepfool}, as illustrated in Figure \ref{fig:A1}, efficiently computes adversarial perturbations to fool the ML model into making incorrect predictions. This illustration strongly signifies the need for the development of an effective detection mechanism to safeguard AI models against this type of adversarial attack.


With the above motivation, for robust AI security analysis, we introduce \textbf{SAMURAI} (\textbf{\underline{S}}afeguarding \textbf{\underline{A}}gainst \textbf{\underline{M}}alicious \textbf{\underline{U}}sage and \textbf{\underline{R}}esilience of \textbf{\underline{AI}}), a comprehensive framework designed to protect AI accelerators against some of the most significant security threats. SAMURAI employs a multilayered low-cost security approach driven by real-time monitoring and anomaly detection techniques. The block diagram of this framework is shown in Figure \ref{fig:A2}.

This framework integrates the AI Performance Counter (APC), special purpose registers similar to the HPC in a processor, into an AI accelerator and an on-chip ML-based analyzer, known as \textit{TANTO}. APC records the operational traces of AI, which are then processed by the \textit{TANTO} aiming to safeguard AI against misuse, ensuring the integrity and security of AI operations. The training process is conducted directly within the ML accelerator, utilizing the APC traces. This on-device training eliminates the need to transfer APC log files (data) to a central server or cloud, thus maintaining data integrity and security. As a result, there is no requirement for online data transfer for training or decision making (on APC data). SAMURAI outperforms traditional software methods; enhances security and regulatory compliance of AI accelerators; and ensures proper use of AI in wide variety of applications, safeguarding AI hardware from unauthorized use and threats. In summary, we make the following major contributions in the proposed work.

\begin{itemize}
\item We identify the deficiencies of current software-based solutions for protecting AI accelerators. To address them, we introduce SAMURAI -- a framework that integrates APC with an on-chip TANTO. SAMURAI leverages APCs to capture low-level hardware events and data patterns at internal nodes specific to AI operations. 

    \item SAMURAI enhances attack resilience by incorporating TANTO (\textbf{\underline{T}}rained \textbf{\underline{A}}nomaly \textbf{\underline{I}}nspection \textbf{\underline{T}}hrough \textbf{\underline{O}}bservation), a lightweight ML model, which performs real-time anomaly detection on AI operations to prevent malicious activities. This method ensures precise monitoring at minimal overhead.

    \item TANTO provides real-time on-device training and inferencing, eliminating the need for data transfer to a central server. This maintains data integrity and ensures efficient detection of adversarial activities in AI applications.

    \item Through extensive experiments on benchmark data and open-source AI models, we demonstrate that SAMURAI supports continuous and adaptive monitoring to enable robust protection against these threats without substantial performance degradation.
\end{itemize}

The remainder of the paper is structured as follows. Section II covers the preliminaries and reviews related work. Section III presents the methodology. Section IV explains the experimental findings and discussions. Finally, Section V presents the conclusions and future research directions.


\section{\textbf{Background and Related works}}
Various attacks on neural network models can be broadly side channel attacks, fault injection attacks, trojan insertion attacks, adversarial input attacks etc, \cite{madry2017towards, moosavi2016deepfool,mittal2021survey,hong2018security, rakin2019bit, fredrikson2015model,liang2022adversarial}.


\textbf{Threat Model:} In this work, the performance of AI architecture is analyzed under both adversarial data generated by the adversarial attack and non-adversarial or clean data. We use an important adversarial attack known as the DeepFool attack \cite{moosavi2016deepfool} for the validation of the proposed scheme. DeepFool attack iteratively pushes the input data across the decision boundary of the classifier, making the perturbation as small as possible to remain imperceptible to humans. An adversarial perturbation is defined as the minimal perturbation \textit{p} that alters the classifier's label $\hat t$(\textit{a}) for a given input as given below.
\begin{equation}
    \Delta(a ;\hat t):= \min_{p} ||p||_{2} \quad subject \quad to \quad \hat t (a+p) \neq \hat t(a)
\end{equation}
where \textit{a} is an image and $\hat t$ (a) is the estimated label. We call $\Delta (a; \hat t)$ the robustness of $\hat t$ at the point \emph{a}. The robustness of classifier $\hat t$ is then defined as
\begin{equation}
    \rho_{adv} (\hat t) = E_{a} \frac{\Delta(a; \hat t)}{||a||_{2}}
\end{equation}
where $E_a$ is the expectation over the distribution of data.

Each type of attack requires different defense mechanisms.


\textbf{AI Accelerator parameters extracted by the APC:} The architectural traces of AI core are captured in real-time by APC. These values are subsequently processed by TANTO unit of SAMURAI. The AI metrics are shown in the table \ref{tab:layer_metrics}.

\begin{table}[h!]
\Large
\centering
\caption{AI accelerator metrics extracted by APC} 
\renewcommand{\arraystretch}{1.2} 
\resizebox{\columnwidth}{!}{ 
\begin{tabular}{|l|p{12cm}|}
\hline
\textbf{Metric} & \textbf{Description and Equation} \\ \hline

\textbf{Layer Sparsity} & Ratio of zero-value elements in a layer's activation. \\
& \( \boldsymbol{S_L = \frac{\#\{o_i \in O | o_i = 0\}}{\#\{o_i \in O\}}} \) \\
&  \( \boldsymbol{o_i} \) is the \( i \)-th element in the output tensor \( \boldsymbol{O} \), and \( \boldsymbol{\#\{\cdot\}} \) denotes the cardinality of the set, \( \boldsymbol{L} \) denotes the neural network layer with output tensor
\(\boldsymbol{O}\), \(\boldsymbol{S_L}\) is the layer sparsity. \\ \hline

\textbf{Layer Zero Counts} & It is the absolute count of zero-
valued elements within each layer’s activation. \\
& \( \boldsymbol{Z_L = \sum_{i=1}^{N} \delta(o_i = 0)} \) \\
&  \( \boldsymbol{N} \) is the total number of elements in \( \boldsymbol{O} \), \( \boldsymbol{o_i} \) is the \( i \)-th element of \( \boldsymbol{O} \), and \( \boldsymbol{\delta(\cdot)} \) is the indicator function. \\ \hline

\textbf{Dense Layer Activities} & Includes average activation, sparsity, and extreme values. \\
& \textbf{Average Activation:} \( \boldsymbol{\bar{a}_L = \frac{1}{N} \sum_{i=1}^{N} \sum_{j=1}^{M} a_{ij}} \) \\
& \textbf{Sparsity:} \( \boldsymbol{S_L = \frac{\#\{a_{ij}\in A | a_{ij} = 0\}}{\#\{a_{ij}\in A\}}} \) \\
& \textbf{Max/Min Activation:} \( \boldsymbol{a_{\max} = \max(A), a_{\min} = \min(A)} \) \\
&  \( \boldsymbol{\bar{a}_L} \) is the average activation, \( \boldsymbol{S_L} \) is the sparsity, \( \boldsymbol{a_{\max}} \) and \( \boldsymbol{a_{\min}} \) are the maximum and minimum activations, \( \boldsymbol{A} \) is the activation matrix, \( \boldsymbol{N} \) is the number of inputs, and \( \boldsymbol{M} \) is the number of neurons in the dense layer. \\ \hline

\textbf{FLOPs per Layer} & Total floating-point computations per layer. \\ \hline

\textbf{Layer-wise TOPs} & Tera Operations Per Second per layer, scaled FLOPs. \\ \hline

\textbf{Layer-wise MACs} & Multiply-Accumulate operations per output element. \\ \hline

\textbf{Layer-wise FLOPs} & Individual FLOPs assessment per layer. \\ \hline

\textbf{Layer-wise Entropies} & Entropy of output distribution. \\
& \( \boldsymbol{H(A) = -\sum_{i=1}^{N} P(a_i) \log P(a_i)} \) \\
&  \( \boldsymbol{H(A)} \) is the entropy of the output distribution, \( \boldsymbol{P(a_i)} \) is the probability of the \( i \)-th outcome, and \( \boldsymbol{N} \) is the total number of classes or distinct outcomes. \\ \hline

\textbf{Throughput} & Inverse of inference time; images processed per second. \\ \hline

\end{tabular}
}
\label{tab:layer_metrics}
\end{table}

\begin{figure*}[!htb]
    \centering
    \includegraphics[width=1\textwidth]{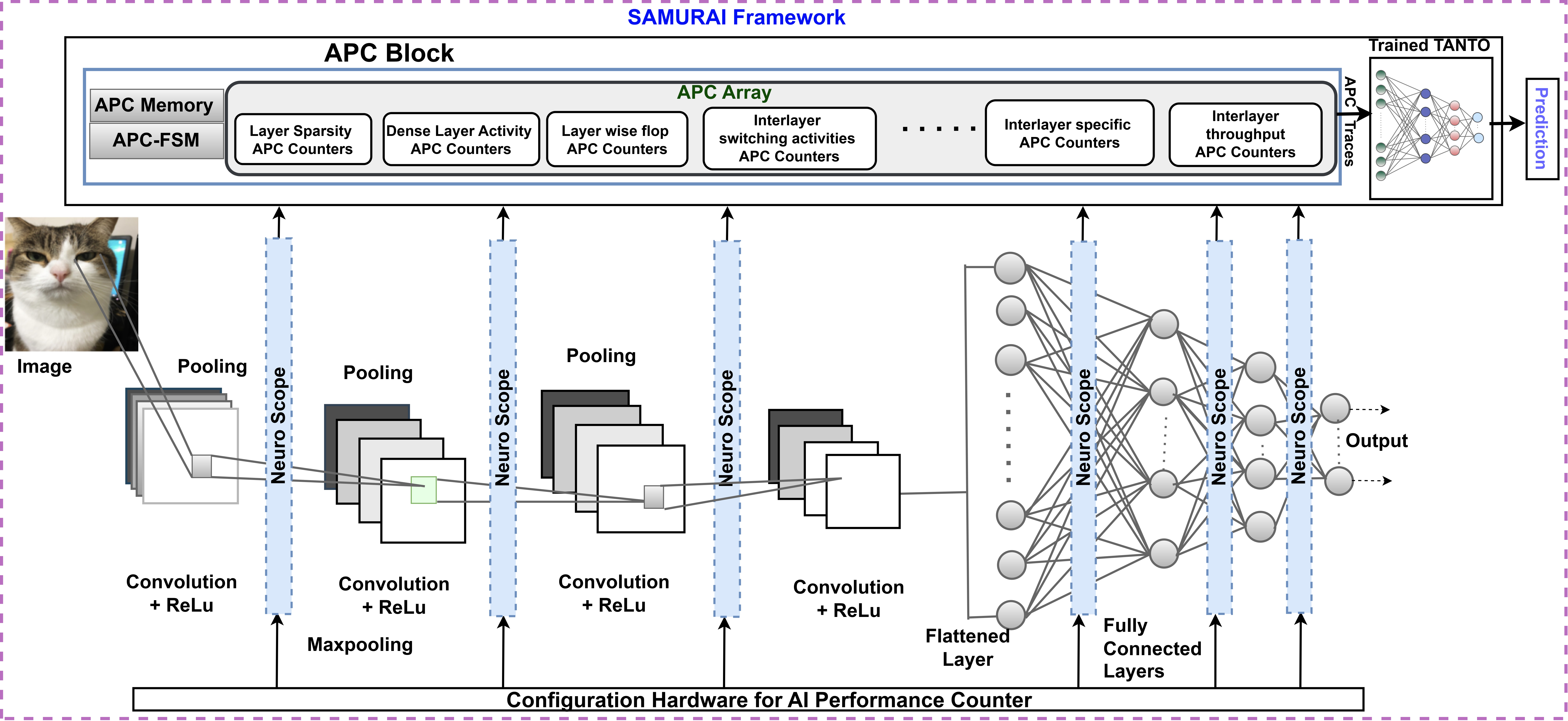}

    \caption {AI Performance Counter (APC) in an AI Accelerator illustrating the architecture of the APC module, which consists of a Finite State Machine (FSM) and a secure APC memory block that records operational metrics during AI inference and securely transmits the data to TANTO for real-time analysis.}
    \label{fig:2}
\end{figure*}

\textbf{Related Works:} The work \cite{moosavi2016deepfool} presented the DeepFool algorithm, which computes minimal perturbations that mislead classifiers, allowing for a reliable robustness assessment. A side channel attack (SCA) using HPCs to infer the structure of DNNs has been reported in \cite{kumar2021inferring}. The work reported in \cite{madry2017towards} highlights the vulnerability of DNNs to adversarial attacks and offers an approach to the detection of attacks on networks. A detection method `Uniguard' proposed in \cite{yan2023unified} safeguards FPGA-based AI accelerators using power side channel information generated during inference to spot anomalies. 
The work \cite{mittal2021survey} highlights the techniques for detecting mainly SCA and Fault Injection attacks in the DNN models and AI Accelerator to enhance the security of AI accelerator. These works highlight the security analysis using software tools. 

Existing software-based solutions target mainly one specific attack and require implementation into the model, making them less practical. The security analysis in AI by examining HPC traces of the processor based on AI operation faces challenges in profiling low-level hardware events of AI. The proposed framework is more integrated, straightforward, and versatile with fewer setup complexity.

\section{\textbf{Methodology}}

In the proposed security framework, the APC meticulously records low-level AI metrics during both the inference and training phases. These metrics are subsequently analyzed by an on-chip \textit{TANTO} in real time to detect and mitigate potential security threats, such as adversarial attacks or unauthorized usage of the AI accelerator. 
This approach not only enhances the security of AI accelerators by providing robust defense mechanisms against malicious activities, but also ensures efficient utilization of computational resources, making it a comprehensive solution for safeguarding AI in critical applications. SAMURAI comprises an APC block and a trained ML Analysis Engine known as TANTO as shown in \ref{fig:2}. Next, we discuss the two basic operations of SAMURAI. 

\subsection{\textbf{AI Performance Counter (APC)}}

AI Performance Counter (APC) is integrated with the AI accelerator, as illustrated in Figure \ref{fig:2}, and is designed to collect a diverse set of operational parameters. The APC block consists of an APC array, a Finite State Machine (FSM) block, and a secure APC memory block. This setup captures the key performance metrics, as detailed in Table \ref{tab:layer_metrics}, during inference.  
To provide granular control over performance monitoring, the \textit{configuration hardware for APC} allows dynamic selection of which counters to activate during execution. This flexibility enables users to tailor the monitoring process based on specific AI workloads, optimizing both resource utilization and security analysis. The configuration hardware facilitates the activation or deactivation of the APC counters to collect only the most relevant performance metrics, minimizing overhead while maximizing insight. Through Neuro Scope, the traces of the neural network are extracted in real time. 
These extracted traces are hooked into the APC block and captured into a set of specialized APCs, collectively known as the \textit{APC Array}. The extracted data are securely stored within the \textit{APC Memory Block}, maintaining both integrity and confidentiality. Furthermore, the APC framework incorporates a \textit{hooking mechanism} that enables real-time interception of the execution of AI traces. This mechanism ensures that critical data points such as layer sparsity, floating-point operations (FLOPs), and memory access patterns—are captured efficiently without disrupting the normal execution flow of the AI model. Once recorded, these performance metrics are transmitted to \textit{SAMURAI’s TANTO module} via a secure communication channel, as shown in Figure \ref{fig:A2}. TANTO processes these traces to detect potential adversarial attacks and predict security threats in real time, ensuring robust protection against AI misuse and vulnerabilities. This dynamic and configurable approach makes the AI Performance Counter a powerful tool to monitor AI accelerator behavior, detect anomalies, and protect AI hardware against adversarial threats.


\begin{figure*}[!h]
    \centering
    \includegraphics[width=1\textwidth]{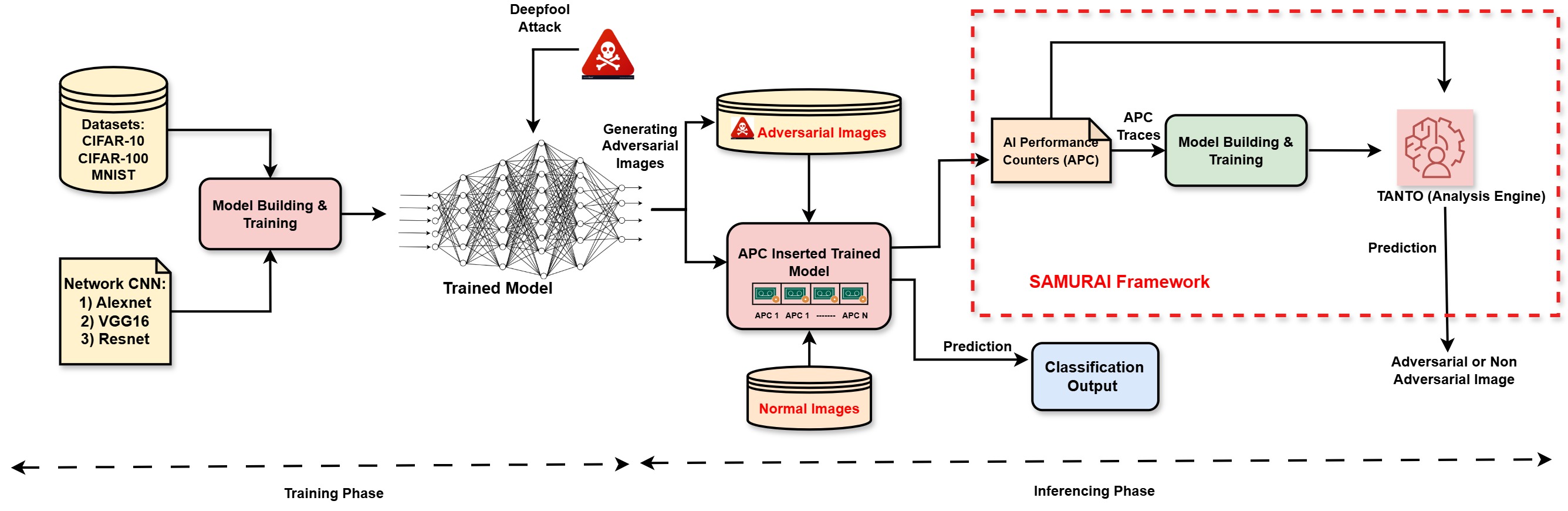}

    \caption{Adversarial Image Detection Process in SAMURAI showing how the SAMURAI framework detects adversarial images by collecting AI Performance Counter (APC) traces, analyzing them in real time, and identifying adversarial inputs during inference.}
    \label{fig:4}
\end{figure*}

\subsection{\textbf{ML Analysis Engine (TANTO) }} 
In this framework, the data collected by the APC is transferred to the ML Engine (\textit{TANTO}). Subsequently, the \textit{TANTO} determines whether the operations executed by the AI accelerator exhibit unauthorized or adversarial intent. It also provides continuous feedback to the AI Accelerator both in the training and inference phase, aimed at curtailing the operations to achieve expedited convergence as well as to effectuate a reduction in power consumption. Basically, \textit{TANTO} monitors the workloads on AI accelerators. The approach identifies and mitigates any potentially malicious activity within the workloads and alerts the user accordingly. Additionally, the AI IP owner can restrict the usability or specific inferencing capabilities of trained models. 


In the next section, we discuss the operations of the proposed scheme.  

\subsection{\textbf{TANTO Training Process}}
The TANTO training process is illustrated in Figure \ref{fig:4}. The AI Accelerator performs inferencing on both clean and adversarial data using trained neural networks. Adversarial images are generated using the \textit{DeepFool attack}. During this process, the AI Performance Counter (APC) captures operational traces  for both normal and adversarial conditions. These traces contain key metrics such as layer sparsity and activation patterns as mentioned in table \ref{tab:layer_metrics}.

To construct the training dataset, APC metrics from both clean and adversarial inputs are combined. This dataset is then used to train the machine learning-based engine, TANTO. The model training phase utilizes standard IMAGE datasets including CIFAR-10, CIFAR-100,and MNIST with widely used CNN models such as ResNet, VGG, and AlexNet. The captured features are used to train TANTO, enabling it to distinguish between original and adversarial input based on APC-derived metrics.

Once trained, TANTO is deployed in edge devices for real-time adversarial attack detection. During inference, it continuously monitors the APC traces and identifies deviations from the expected behavior. When adversarial patterns are detected, TANTO triggers security measures such as issuing user alerts or stopping compromised operations, ensuring the integrity of AI systems. The following algorithms describe the process flow step by step.

\begin{algorithm}
\small
\caption{Train\_Model}
\begin{algorithmic}[1]
\Procedure{Train\_Model}{}
    \State \textbf{Input:} Training dataset $\mathcal{D}_{\text{train}} = \{(\mathbf{x}_i, y_i)\}_{i=1}^{N}$
    \State \textbf{Initialize:} Neural network $M$ with parameters $\theta$, 
    \Statex \hspace{15pt} loss function $\mathcal{L}$, learning rate $\eta$, and optimizer $\text{opt}$
    \For{$\text{epoch} = 1$ \textbf{to} $E$}
        \For{each $(\mathbf{x},y) \in \mathcal{D}_{\text{train}}$}
            \State $\hat{y} = M(\mathbf{x}; \theta)$
            \State $J = \mathcal{L}(\hat{y}, y)$
            \State $\theta \leftarrow \theta - \eta \nabla_{\theta} J$
        \EndFor
    \EndFor
    \State \textbf{Output:} Trained model $M_{\text{trained}}$
\EndProcedure
\end{algorithmic}
\end{algorithm}

The \textbf{Algorithm 1} involves training a neural network model using a standard training dataset $\mathcal{D}_{\text{train}} = \{(\mathbf{x}_i, y_i)\}_{i=1}^{N}$
,where $\mathbf{x}_i$ represents the input images and $y_i$ represents the corresponding labels. The model $M$ is initialized with parameters $\theta$, a loss function $\mathcal{L}$, a learning rate $\eta$, and an optimizer $\text{opt}$. During training, the model parameters are iteratively updated to minimize the loss function over multiple epochs $E$.

Once the model is trained, then \textbf{Algorithm 2} involves extracting APC metrics from the test dataset $\{\mathbf{APC}_{\text{original}}(\mathbf{x}_i)\}_{i=1}^{M}$. Each test image $\mathbf{x}$ is passed through the trained model $M_{\text{trained}}$, and internal activations (APC metrics) are captured using hook functions. These metrics provide a detailed representation of the model's internal state and are crucial for detecting adversarial perturbations. These images are marked as non-adversarial, forming the set of original APC metrics 
$\{\mathbf{APC}_{\text{original}}(\mathbf{x}_i)\}_{i=1}^{M}$.


\begin{algorithm}
\small
\caption{Extract\_APC\_Metrics}
\begin{algorithmic}[1]
\Procedure{Extract\_APC\_Metrics}{}
    \State \textbf{Input:} Trained model $M_{\text{trained}}$, test dataset $\mathcal{D}_{\text{test}} = \{\mathbf{x}_i\}_{i=1}^{M}$
    \For{each $\mathbf{x} \in \mathcal{D}_{\text{test}}$}
        \State $\hat{y} = M_{\text{trained}}(\mathbf{x})$
        \State Extract APC metrics: $\mathbf{APC}_{\text{original}}(\mathbf{x}) = \text{hook}(M_{\text{trained}}, \mathbf{x})$
        \State Mark image $\mathbf{x}$ as non-adversarial
    \EndFor
    \State \textbf{Output:} APC metrics $\{\mathbf{APC}_{\text{original}}(\mathbf{x}_i)\}_{i=1}^{M}$
\EndProcedure
\end{algorithmic}
\end{algorithm}


\begin{algorithm}
\small
\caption{DeepFool\_Attack}
\begin{algorithmic}[1]
\Procedure{DeepFool\_Attack}{}
    \State \textbf{Input:} Trained model $M_{\text{trained}}$, test dataset $\mathcal{D}_{\text{test}} = \{\mathbf{x}_i\}_{i=1}^{M}$
    \For{each $\mathbf{x} \in \mathcal{D}_{\text{test}}$}
        \State Generate adversarial example: 
        \Statex \hspace{29pt}$\mathbf{x}_{\text{adv}} = \text{DeepFool}(M_{\text{trained}}, \mathbf{x})$
        \State $\hat{y}_{\text{adv}} = M_{\text{trained}}(\mathbf{x}_{\text{adv}})$
        \State Extract APC metrics: 
        \Statex \hspace{25pt}$\mathbf{APC}_{\text{adv}}(\mathbf{x}_{\text{adv}}) = \text{hook}(M_{\text{trained}}, \mathbf{x}_{\text{adv}})$
    \EndFor
    \State \textbf{Output:} APC metrics $\{\mathbf{APC}_{\text{adv}}(\mathbf{x}_{\text{adv},i})\}_{i=1}^{M}$
\EndProcedure
\end{algorithmic}
\end{algorithm}


Next, in \textbf{Algorithm 3}, the adversarial perturbations are generated from the test data set using the DeepFool attack, which iteratively perturbs the input images $\mathbf{x}$ to create adversarial images $\mathbf{x}_{\text{adv}}$. These adversarial images are then processed through the trained model $M_{\text{trained}}$ to extract their APC metrics. This results in a set of adversarial APC metrics $\{\mathbf{APC}_{\text{adv}}(\mathbf{x}_{\text{adv},i})\}_{i=1}^{M}$.

The \textbf{Algorithm 4} shows that the original and adversarial APC metrics are combined to build a new dataset $\mathcal{D}_{\text{APC}} = \{(\mathbf{APC}_{\text{original},i}, 0), (\mathbf{APC}_{\text{adv},i}, 1)\}_{i=1}^{M}$. This data set is used to train a machine learning-based detector model $D_{\text{ML}}$, which learns to distinguish between original and adversarial images based on APC metrics.


\begin{algorithm}
\small
\caption{Train\_Detector\_Model}
\begin{algorithmic}[1]
\Procedure{Train\_Detector\_Model}{}
    \State \textbf{Input:} APC metrics from original and adversarial images \Statex \hspace{45pt}$\{\mathbf{APC}_{\text{original}}, \mathbf{APC}_{\text{adv}}\}$
    \State Construct dataset: 
    \[
    \mathcal{D}_{\text{APC}} = \{(\mathbf{APC}_{\text{original},i}, 0), (\mathbf{APC}_{\text{adv},i}, 1)\}_{i=1}^{M}
    \]
    \State Train machine learning model $D_{\text{ML}}$ on $\mathcal{D}_{\text{APC}}$
    \State \textbf{Output:} Trained detector model (TANTO) $D_{\text{ML,trained}}$
\EndProcedure
\end{algorithmic}
\end{algorithm}


\subsection{\textbf{Monitoring and Attack Detection}}
Figure \ref{fig:4} illustrates the adversarial input detection methodology within the SAMURAI framework, describing the workflow designed to safeguard AI systems against adversarial attacks. 
The APCs of AI Accelerator are combined with trained TANTO, forming the SAMURAI framework which is integrated with the AI Accelerators, as shown in Figure \ref{fig:4}. The process uses image input data sets such as CIFAR-10, CIFAR-100, MNIST and trained neural network (CNN) models such as ResNet, VGG, and AlexNet. 
The figure provides a step-by-step depiction of the detection process, ensuring robust security and resilience in AI-driven applications.

During inference of AI Accelerator using unlabeled adversarial or nonadversarial Image input, APCs track and record key AI operations metrics of AI operations as shown in Table \ref{tab:layer_metrics} to monitor model behavior under both clean and adversarial conditions. APC generates secure APC tracers for further analysis, which TANTO examines to identify adversarial patterns by comparing operational differences between clean and adversarial data. During inference of TANTO, TANTO continuously monitors APC traces to detect deviations from expected behavior, enabling real-time adversarial detection. Upon detection, the system triggers security responses, such as alerts or stopping operations. Figure \ref{fig:4} illustrates this integrated approach using hardware-level monitoring and machine learning to improve AI security against adversarial threats.

By analyzing the traces during the process, TANTO can detect whether the image presented to the model is adversarial or not. The proposed framework for adversarial image detection and classification involves a multistage process aimed at enhancing the robustness of neural network models against adversarial attacks. Algorithm 5 describes the process flow for attack detection.

\begin{algorithm}
\small
\caption{Infer\_And\_Detect}
\begin{algorithmic}[1]
\Procedure{Infer\_And\_Detect}{}
    \State \textbf{Input:} Trained model $M_{\text{trained}}$, trained detector model 
    \Statex \hspace{39pt}$D_{\text{ML,trained}}$, workload dataset $\mathcal{D}_{\text{workload}}$
    \For{each $\mathbf{x}_{\text{workload}} \in \mathcal{D}_{\text{workload}}$}
        \State $\hat{y}_{\text{workload}} = M_{\text{trained}}(\mathbf{x}_{\text{workload}})$
        \State Extract APC metrics: 
        \Statex \hspace{25pt}$\mathbf{APC}_{\text{workload}}(\mathbf{x}_{\text{workload}}) = \text{hook}(M_{\text{trained}}, \mathbf{x}_{\text{workload}})$
        \State Classify $\mathbf{x}_{\text{workload}}$ using $D_{\text{ML,trained}}$:
        \[
        \text{is\_adv} = D_{\text{ML,trained}}(\mathbf{APC}_{\text{workload}})
        \]
    \EndFor
    \State \textbf{Output:} Classification results for workload dataset
\EndProcedure
\end{algorithmic}
\end{algorithm}

The \textbf{Algorithm 5} shows that during the inference phase, new images of a workload dataset $\mathcal{D}_{\text{workload}}$ are processed through the trained model $M_{\text{trained}}$ to extract APC metrics. These metrics are then classified by the trained detector model $D_{\text{ML,trained}}$ to determine if the images are adversarial. The detector model outputs a classification that indicates whether each image is adversarial or not. This comprehensive approach enhances the security and robustness of neural network models against adversarial attacks by leveraging the detailed internal activations captured in the APC metrics.

In the experiment section, the results are analyzed to verify the performance of the AI accelerator.

\begin{figure*}[] 
    \centering
    \begin{subfigure}{0.3\textwidth} 
       \includegraphics[width=\linewidth]{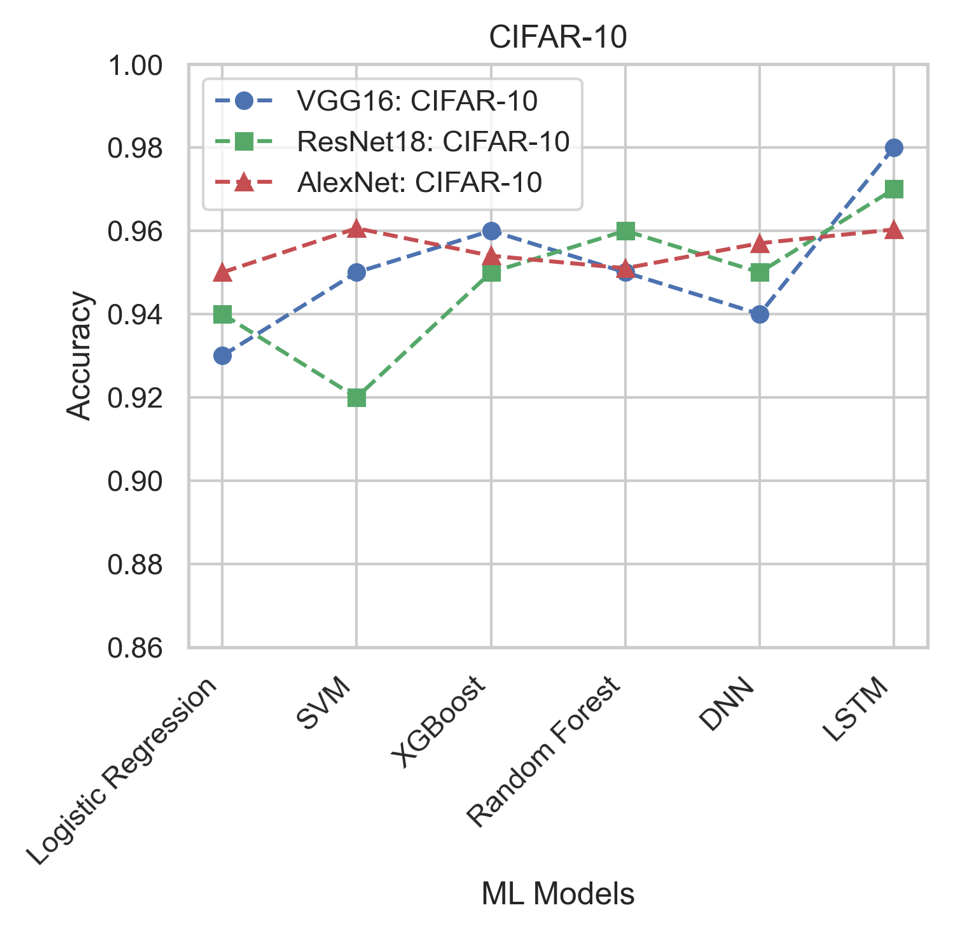}
        \caption{}
        \label{fig5a} 
    \end{subfigure}
\hfill
    \begin{subfigure}{0.3\textwidth}
        \includegraphics[width=\linewidth]{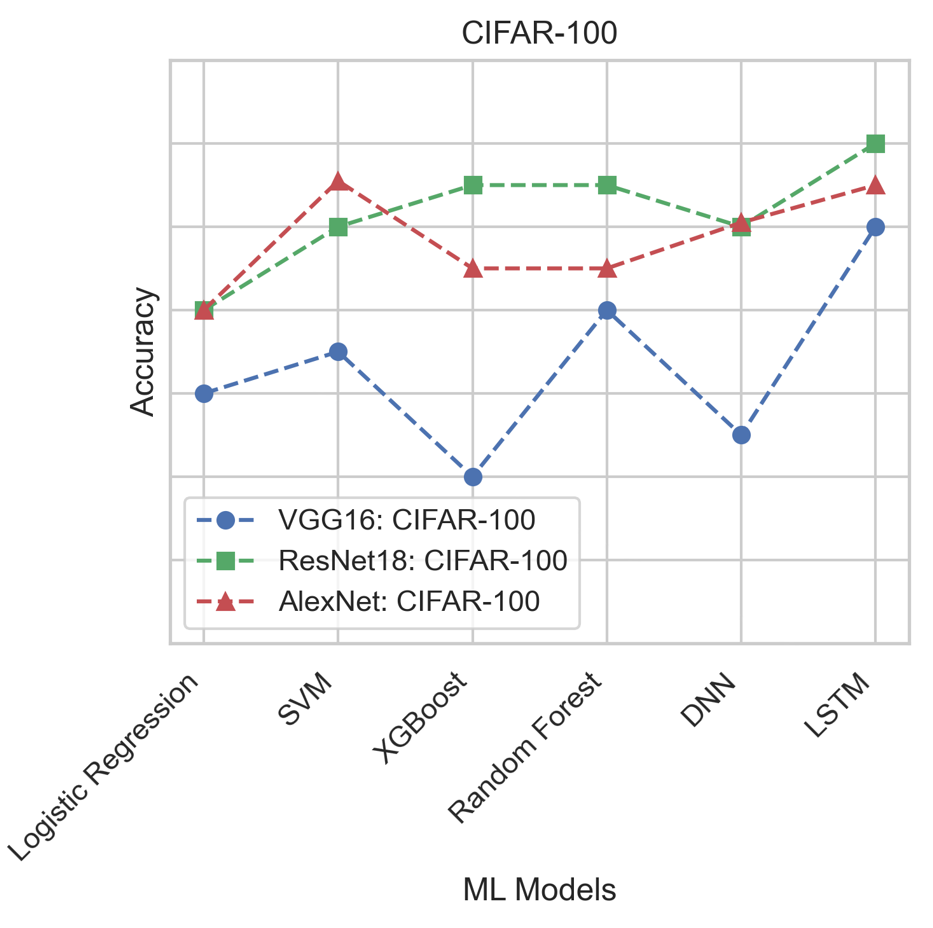}
        \caption{}
        \label{fig5b}
    \end{subfigure}
 \hfill  
    \begin{subfigure}{0.3\textwidth}
        \includegraphics[width=\linewidth]{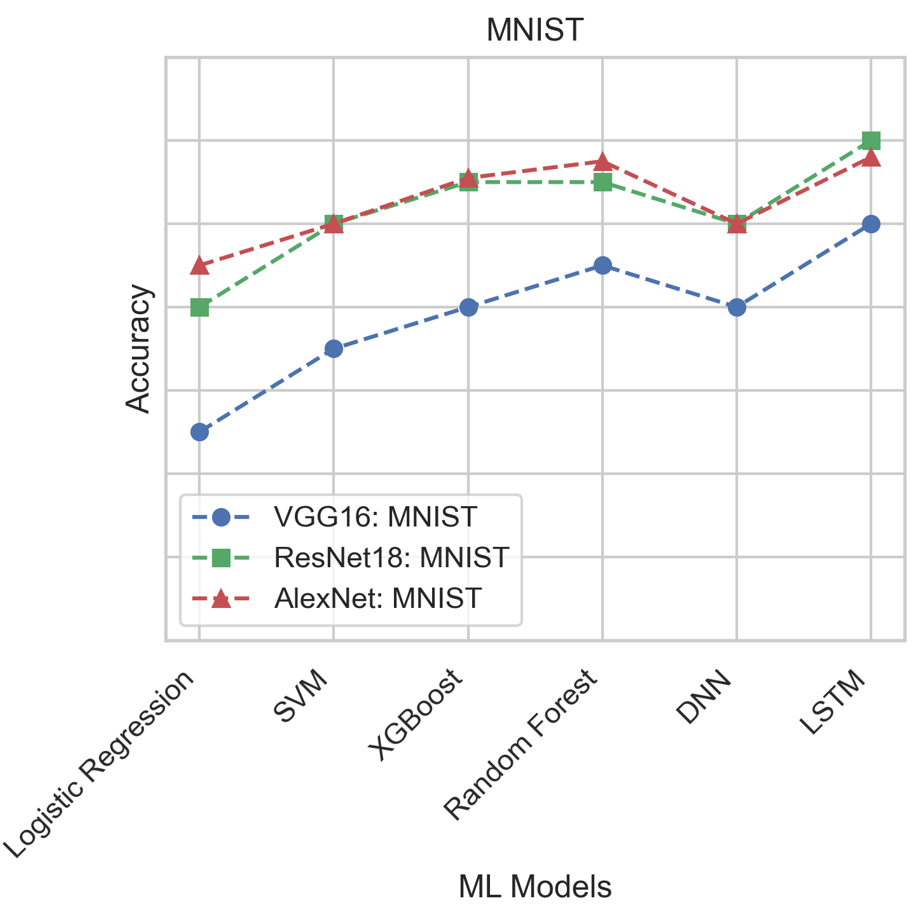}
        \caption{}
        \label{fig5c}
    \end{subfigure}

     \caption{Comparison of Different ML Models for Adversarial Image Detection presenting the accuracy of various machine learning models, including Logistic Regression, SVM, XGBoost, Random Forest, DNN, and LSTM, in distinguishing between adversarial and non-adversarial inputs.}
    \label{fig5}
    \end{figure*}
    
\section{\textbf{Experimental Results and Discussions}}

We use DeepFool Attack \cite{moosavi2016deepfool} to validate the proposed technique. 







\textbf{Experimental Inputs:} \\
The neural networks, datasets, and models used in the works are given below.

\begin{itemize}
    \item \textbf{Neural Network:} \\
    (a) Vgg16, (b) Resnet18, (c) Alexnet
    \item \textbf{Dataset:} \\
    (a) CIFAR-10, (b) CIFAR-100, (c) MNIST
    \item \textbf{ML Models used:} \\
    Logistic Regression, SVM, XGBoost, Random Forest, DNN, and LSTM
\end{itemize}

During the inference process, the APC traces for both adversarial and non-adversarial instances are recorded and are subsequently used to train the \textit{TANTO}. Following training, \textit{TANTO} determines whether an image input to the AI accelerator is adversarial or not. The flow of the ML model with the APC parameter to detect anomalies in the data is shown in Figure \ref{fig:4}.

The results of the Deepfool attack performed on different neural network architectures based on different data sets are presented in Table \ref{tab:my-table_4}. Table \ref{tab:my-table_4} shows the performance metrics of the Logistic Regression, SVM, XGBoost, Random Forest, DNN,and LSTM models in detecting adversarial perturbations, and all ML models achieve significant detection accuracy (more than 90\%). Table \ref{tab:my-table_4}  indicates that the LSTM model consistently achieves the highest accuracy (0.98 for CIFAR-10 and ResNet18) establishing its robustness over the different data distributions and complexities. These findings highlight that LSTM models are particularly effective in utilizing APC metrics for adversarial detection, due to their ability to capture.

\begin{table}[h]
\centering
\caption{\textbf{\scriptsize
Comparison of different ML Models for Prediction of Adversarial or Non-Adversarial images }}
\label{tab:my-table_4}
\resizebox{\columnwidth}{!}{%
\begin{tabular}{|
>{\columncolor[HTML]{FFFFFF}}c |
>{\columncolor[HTML]{FFFFFF}}c |
>{\columncolor[HTML]{FFFFFF}}c |
>{\columncolor[HTML]{FFFFFF}}c 
>{\columncolor[HTML]{FFFFFF}}c 
>{\columncolor[HTML]{FFFFFF}}c 
>{\columncolor[HTML]{FFFFFF}}c 
>{\columncolor[HTML]{FFFFFF}}c 
>{\columncolor[HTML]{FFFFFF}}c |}
\hline
\cellcolor[HTML]{FFFFFF} &
  \cellcolor[HTML]{FFFFFF} &
  \cellcolor[HTML]{FFFFFF} &
  \multicolumn{6}{c|}{\cellcolor[HTML]{FFFFFF}\textbf{\begin{tabular}[c]{@{}c@{}}Comparison of different ML Models for   \\ Prediction of Adversarial or Non-Adversarial\end{tabular}}} \\ \cline{4-9} 
\multirow{-2}{*}{\cellcolor[HTML]{FFFFFF}\textbf{Network}} &
  \multirow{-2}{*}{\cellcolor[HTML]{FFFFFF}\textbf{Dataset}} &
  \multirow{-2}{*}{\cellcolor[HTML]{FFFFFF}\textbf{\begin{tabular}[c]{@{}c@{}}APC Model Traces \\   file in (MB)\end{tabular}}} &
  \multicolumn{1}{c|}{\cellcolor[HTML]{FFFFFF}\textbf{\begin{tabular}[c]{@{}c@{}}Logistic \\ Regression\end{tabular}}} &
  \multicolumn{1}{c|}{\cellcolor[HTML]{FFFFFF}\textbf{SVM}} &
  \multicolumn{1}{c|}{\cellcolor[HTML]{FFFFFF}\textbf{XGBoost}} &
  \multicolumn{1}{c|}{\cellcolor[HTML]{FFFFFF}\textbf{\begin{tabular}[c]{@{}c@{}}Random \\ Forest\end{tabular}}} &
  \multicolumn{1}{c|}{\cellcolor[HTML]{FFFFFF}\textbf{DNN}} &
  \textbf{LSTM} \\ \hline
\cellcolor[HTML]{FFFFFF} &
  \textbf{CIFAR-10} &
  280 MB &
  \multicolumn{1}{c|}{\cellcolor[HTML]{FFFFFF}0.93} &
  \multicolumn{1}{c|}{\cellcolor[HTML]{FFFFFF}0.95} &
  \multicolumn{1}{c|}{\cellcolor[HTML]{FFFFFF}0.96} &
  \multicolumn{1}{c|}{\cellcolor[HTML]{FFFFFF}0.95} &
  \multicolumn{1}{c|}{\cellcolor[HTML]{FFFFFF}0.94} &
  0.98 \\ \cline{2-9} 
\cellcolor[HTML]{FFFFFF} &
  \cellcolor[HTML]{FFFFFF} &
  \cellcolor[HTML]{FFFFFF} &
  \multicolumn{1}{c|}{\cellcolor[HTML]{FFFFFF}} &
  \multicolumn{1}{c|}{\cellcolor[HTML]{FFFFFF}} &
  \multicolumn{1}{c|}{\cellcolor[HTML]{FFFFFF}} &
  \multicolumn{1}{c|}{\cellcolor[HTML]{FFFFFF}} &
  \multicolumn{1}{c|}{\cellcolor[HTML]{FFFFFF}} &
  \cellcolor[HTML]{FFFFFF} \\
\cellcolor[HTML]{FFFFFF} &
  \cellcolor[HTML]{FFFFFF} &
  \cellcolor[HTML]{FFFFFF} &
  \multicolumn{1}{c|}{\cellcolor[HTML]{FFFFFF}} &
  \multicolumn{1}{c|}{\cellcolor[HTML]{FFFFFF}} &
  \multicolumn{1}{c|}{\cellcolor[HTML]{FFFFFF}} &
  \multicolumn{1}{c|}{\cellcolor[HTML]{FFFFFF}} &
  \multicolumn{1}{c|}{\cellcolor[HTML]{FFFFFF}} &
  \cellcolor[HTML]{FFFFFF} \\
\cellcolor[HTML]{FFFFFF} &
  \multirow{-3}{*}{\cellcolor[HTML]{FFFFFF}\textbf{CIFAR-100}} &
  \multirow{-3}{*}{\cellcolor[HTML]{FFFFFF}300 MB} &
  \multicolumn{1}{c|}{\multirow{-3}{*}{\cellcolor[HTML]{FFFFFF}0.92}} &
  \multicolumn{1}{c|}{\multirow{-3}{*}{\cellcolor[HTML]{FFFFFF}0.93}} &
  \multicolumn{1}{c|}{\multirow{-3}{*}{\cellcolor[HTML]{FFFFFF}0.9}} &
  \multicolumn{1}{c|}{\multirow{-3}{*}{\cellcolor[HTML]{FFFFFF}0.94}} &
  \multicolumn{1}{c|}{\multirow{-3}{*}{\cellcolor[HTML]{FFFFFF}0.91}} &
  \multirow{-3}{*}{\cellcolor[HTML]{FFFFFF}0.96} \\ \cline{2-9} 
\cellcolor[HTML]{FFFFFF} &
  \cellcolor[HTML]{FFFFFF} &
  \cellcolor[HTML]{FFFFFF} &
  \multicolumn{1}{c|}{\cellcolor[HTML]{FFFFFF}} &
  \multicolumn{1}{c|}{\cellcolor[HTML]{FFFFFF}} &
  \multicolumn{1}{c|}{\cellcolor[HTML]{FFFFFF}} &
  \multicolumn{1}{c|}{\cellcolor[HTML]{FFFFFF}} &
  \multicolumn{1}{c|}{\cellcolor[HTML]{FFFFFF}} &
  \cellcolor[HTML]{FFFFFF} \\
\multirow{-6}{*}{\cellcolor[HTML]{FFFFFF}\textbf{Vgg16}} &
  \multirow{-2}{*}{\cellcolor[HTML]{FFFFFF}\textbf{MNIST}} &
  \multirow{-2}{*}{\cellcolor[HTML]{FFFFFF}119.8MB} &
  \multicolumn{1}{c|}{\multirow{-2}{*}{\cellcolor[HTML]{FFFFFF}0.91}} &
  \multicolumn{1}{c|}{\multirow{-2}{*}{\cellcolor[HTML]{FFFFFF}0.93}} &
  \multicolumn{1}{c|}{\multirow{-2}{*}{\cellcolor[HTML]{FFFFFF}0.94}} &
  \multicolumn{1}{c|}{\multirow{-2}{*}{\cellcolor[HTML]{FFFFFF}0.95}} &
  \multicolumn{1}{c|}{\multirow{-2}{*}{\cellcolor[HTML]{FFFFFF}0.94}} &
  \multirow{-2}{*}{\cellcolor[HTML]{FFFFFF}0.96} \\ \hline
\cellcolor[HTML]{FFFFFF} &
  \textbf{CIFAR-10} &
  305 MB &
  \multicolumn{1}{c|}{\cellcolor[HTML]{FFFFFF}0.94} &
  \multicolumn{1}{c|}{\cellcolor[HTML]{FFFFFF}0.92} &
  \multicolumn{1}{c|}{\cellcolor[HTML]{FFFFFF}0.95} &
  \multicolumn{1}{c|}{\cellcolor[HTML]{FFFFFF}0.96} &
  \multicolumn{1}{c|}{\cellcolor[HTML]{FFFFFF}0.95} &
  0.97 \\ \cline{2-9} 
\cellcolor[HTML]{FFFFFF} &
  \cellcolor[HTML]{FFFFFF} &
  \cellcolor[HTML]{FFFFFF} &
  \multicolumn{1}{c|}{\cellcolor[HTML]{FFFFFF}} &
  \multicolumn{1}{c|}{\cellcolor[HTML]{FFFFFF}} &
  \multicolumn{1}{c|}{\cellcolor[HTML]{FFFFFF}} &
  \multicolumn{1}{c|}{\cellcolor[HTML]{FFFFFF}} &
  \multicolumn{1}{c|}{\cellcolor[HTML]{FFFFFF}} &
  \cellcolor[HTML]{FFFFFF} \\
\cellcolor[HTML]{FFFFFF} &
  \multirow{-2}{*}{\cellcolor[HTML]{FFFFFF}\textbf{CIFAR-100}} &
  \multirow{-2}{*}{\cellcolor[HTML]{FFFFFF}310MB} &
  \multicolumn{1}{c|}{\multirow{-2}{*}{\cellcolor[HTML]{FFFFFF}0.94}} &
  \multicolumn{1}{c|}{\multirow{-2}{*}{\cellcolor[HTML]{FFFFFF}0.96}} &
  \multicolumn{1}{c|}{\multirow{-2}{*}{\cellcolor[HTML]{FFFFFF}0.97}} &
  \multicolumn{1}{c|}{\multirow{-2}{*}{\cellcolor[HTML]{FFFFFF}0.96}} &
  \multicolumn{1}{c|}{\multirow{-2}{*}{\cellcolor[HTML]{FFFFFF}0.95}} &
  \multirow{-2}{*}{\cellcolor[HTML]{FFFFFF}0.98} \\ \cline{2-9} 
\cellcolor[HTML]{FFFFFF} &
  \cellcolor[HTML]{FFFFFF} &
  \cellcolor[HTML]{FFFFFF} &
  \multicolumn{1}{c|}{\cellcolor[HTML]{FFFFFF}} &
  \multicolumn{1}{c|}{\cellcolor[HTML]{FFFFFF}} &
  \multicolumn{1}{c|}{\cellcolor[HTML]{FFFFFF}} &
  \multicolumn{1}{c|}{\cellcolor[HTML]{FFFFFF}} &
  \multicolumn{1}{c|}{\cellcolor[HTML]{FFFFFF}} &
  \cellcolor[HTML]{FFFFFF} \\
\multirow{-5}{*}{\cellcolor[HTML]{FFFFFF}\textbf{Resnet18}} &
  \multirow{-2}{*}{\cellcolor[HTML]{FFFFFF}\textbf{MNIST}} &
  \multirow{-2}{*}{\cellcolor[HTML]{FFFFFF}517   MB} &
  \multicolumn{1}{c|}{\multirow{-2}{*}{\cellcolor[HTML]{FFFFFF}0.94}} &
  \multicolumn{1}{c|}{\multirow{-2}{*}{\cellcolor[HTML]{FFFFFF}0.96}} &
  \multicolumn{1}{c|}{\multirow{-2}{*}{\cellcolor[HTML]{FFFFFF}0.97}} &
  \multicolumn{1}{c|}{\multirow{-2}{*}{\cellcolor[HTML]{FFFFFF}0.97}} &
  \multicolumn{1}{c|}{\multirow{-2}{*}{\cellcolor[HTML]{FFFFFF}0.96}} &
  \multirow{-2}{*}{\cellcolor[HTML]{FFFFFF}0.98} \\ \hline
\cellcolor[HTML]{FFFFFF} &
  \textbf{CIFAR-10} &
  106.6 MB &
  \multicolumn{1}{c|}{\cellcolor[HTML]{FFFFFF}0.95} &
  \multicolumn{1}{c|}{\cellcolor[HTML]{FFFFFF}0.9606} &
  \multicolumn{1}{c|}{\cellcolor[HTML]{FFFFFF}0.954} &
  \multicolumn{1}{c|}{\cellcolor[HTML]{FFFFFF}0.951} &
  \multicolumn{1}{c|}{\cellcolor[HTML]{FFFFFF}0.957} &
  0.9603 \\ \cline{2-9} 
\cellcolor[HTML]{FFFFFF} &
  \cellcolor[HTML]{FFFFFF} &
  \cellcolor[HTML]{FFFFFF} &
  \multicolumn{1}{c|}{\cellcolor[HTML]{FFFFFF}} &
  \multicolumn{1}{c|}{\cellcolor[HTML]{FFFFFF}} &
  \multicolumn{1}{c|}{\cellcolor[HTML]{FFFFFF}} &
  \multicolumn{1}{c|}{\cellcolor[HTML]{FFFFFF}} &
  \multicolumn{1}{c|}{\cellcolor[HTML]{FFFFFF}} &
  \cellcolor[HTML]{FFFFFF} \\
\cellcolor[HTML]{FFFFFF} &
  \multirow{-2}{*}{\cellcolor[HTML]{FFFFFF}\textbf{CIFAR-100}} &
  \multirow{-2}{*}{\cellcolor[HTML]{FFFFFF}106.6   MB} &
  \multicolumn{1}{c|}{\multirow{-2}{*}{\cellcolor[HTML]{FFFFFF}0.94}} &
  \multicolumn{1}{c|}{\multirow{-2}{*}{\cellcolor[HTML]{FFFFFF}0.971}} &
  \multicolumn{1}{c|}{\multirow{-2}{*}{\cellcolor[HTML]{FFFFFF}0.95}} &
  \multicolumn{1}{c|}{\multirow{-2}{*}{\cellcolor[HTML]{FFFFFF}0.95}} &
  \multicolumn{1}{c|}{\multirow{-2}{*}{\cellcolor[HTML]{FFFFFF}0.961}} &
  \multirow{-2}{*}{\cellcolor[HTML]{FFFFFF}0.97} \\ \cline{2-9} 
\cellcolor[HTML]{FFFFFF} &
  \cellcolor[HTML]{FFFFFF} &
  \cellcolor[HTML]{FFFFFF} &
  \multicolumn{1}{c|}{\cellcolor[HTML]{FFFFFF}} &
  \multicolumn{1}{c|}{\cellcolor[HTML]{FFFFFF}} &
  \multicolumn{1}{c|}{\cellcolor[HTML]{FFFFFF}} &
  \multicolumn{1}{c|}{\cellcolor[HTML]{FFFFFF}} &
  \multicolumn{1}{c|}{\cellcolor[HTML]{FFFFFF}} &
  \cellcolor[HTML]{FFFFFF} \\
\cellcolor[HTML]{FFFFFF} &
  \cellcolor[HTML]{FFFFFF} &
  \cellcolor[HTML]{FFFFFF} &
  \multicolumn{1}{c|}{\cellcolor[HTML]{FFFFFF}} &
  \multicolumn{1}{c|}{\cellcolor[HTML]{FFFFFF}} &
  \multicolumn{1}{c|}{\cellcolor[HTML]{FFFFFF}} &
  \multicolumn{1}{c|}{\cellcolor[HTML]{FFFFFF}} &
  \multicolumn{1}{c|}{\cellcolor[HTML]{FFFFFF}} &
  \cellcolor[HTML]{FFFFFF} \\
\multirow{-6}{*}{\cellcolor[HTML]{FFFFFF}\textbf{Alexnet}} &
  \multirow{-3}{*}{\cellcolor[HTML]{FFFFFF}\textbf{MNIST}} &
  \multirow{-3}{*}{\cellcolor[HTML]{FFFFFF}179 MB} &
  \multicolumn{1}{c|}{\multirow{-3}{*}{\cellcolor[HTML]{FFFFFF}0.95}} &
  \multicolumn{1}{c|}{\multirow{-3}{*}{\cellcolor[HTML]{FFFFFF}0.96}} &
  \multicolumn{1}{c|}{\multirow{-3}{*}{\cellcolor[HTML]{FFFFFF}0.971}} &
  \multicolumn{1}{c|}{\multirow{-3}{*}{\cellcolor[HTML]{FFFFFF}0.975}} &
  \multicolumn{1}{c|}{\multirow{-3}{*}{\cellcolor[HTML]{FFFFFF}0.96}} &
  \multirow{-3}{*}{\cellcolor[HTML]{FFFFFF}0.976} \\ \hline
\end{tabular}%
}
\end{table}

Figure \ref{fig5} compares the performance of various models in detecting adversarial inputs using APC metrics. The graph shows that the LSTM models consistently achieve the highest accuracy in different datasets and different neural network architectures. For example, the LSTM model achieves an accuracy of 0.98 in CIFAR-10 with ResNet18, indicating its robustness in handling diverse data distributions. The figure highlights the effectiveness of LSTM models in utilizing APC metrics for adversarial detection, underscoring their ability to capture the intricate patterns within the data that contribute to their superior performance.

\begin{table}[h]
\centering
\caption{Overhead for combined APC and TANTO over various models and datasets}
\label{tab:my-table-11}
\resizebox{\columnwidth}{!}{%
\begin{tabular}{|l|c|c|c|c|c|}
\hline
\textbf{\begin{tabular}[c]{@{}c@{}}Model \\ (Architecture)\end{tabular}} &
  \textbf{Dataset} &
  \textbf{\begin{tabular}[c]{@{}c@{}}Inference Time \\ without APC(s)\\  (Single \\ Image) (s)\end{tabular}} &
  \textbf{\begin{tabular}[c]{@{}c@{}}TANTO \\ (ML Model) (S)\end{tabular}} &
  \textbf{\begin{tabular}[c]{@{}c@{}}Inference Time for \\ cobmined APC and \\ TANTO Integration (s)\end{tabular}} &
  \textbf{Overhead} \\ \hline
ResNet-34  & CIFAR-10  & 1.05 & 0.219 & 1.269 & 17.30\% \\ \hline
ResNet-18  & CIFAR-100 & 0.8  & 0.056 & 0.856 & 6.50\%  \\ \hline
ResNet-18  & CIFAR-10  & 0.75 & 0.063 & 0.813 & 7.70\%  \\ \hline
AlexNet    & CIFAR-100 & 0.62 & 0.069 & 0.689 & 10.00\% \\ \hline
AlexNet    & CIFAR-10  & 0.58 & 0.034 & 0.614 & 5.50\%  \\ \hline
AlexNet    & MNIST     & 0.52 & 0.046 & 0.566 & 8.10\%  \\ \hline
VGG-16     & CIFAR-10  & 1.25 & 0.165 & 1.415 & 11.70\% \\ \hline
VGG-16     & CIFAR-100 & 1.27 & 0.219 & 1.489 & 14.70\% \\ \hline
VGG-16     & MNIST     & 1.1  & 0.045 & 1.145 & 3.90\%  \\ \hline
ResNet-101 & CIFAR-100 & 1.4  & 0.365 & 1.765 & 20.70\% \\ \hline
ResNet-101 & CIFAR-10  & 1.35 & 0.342 & 1.692 & 20.20\% \\ \hline
ResNet-50  & CIFAR-10  & 1.2  & 0.279 & 1.479 & 18.90\% \\ \hline
ResNet-50  & CIFAR-100 & 1.17 & 0.289 & 1.459 & 19.80\% \\ \hline
\end{tabular}%
}
\end{table}

Table \ref{tab:my-table-11} outlines the computational data on overhead and inference time for different models and datasets with and without the APC integration. For example, 
the AlexNet on MNIST shows an 8.10\% overhead, VGG-16 on CIFAR-100 experiences a 14.70\% overhead, and the ResNet-34 on the CIFAR-10 dataset shows an overhead of 17.30\%, with increase in inference time from 1.05 seconds to 1.269 seconds when APC is integrated. These values vary over the different models. This shows the trade-off between enhanced security and computational efficiency, demonstrating the feasibility of integrating APC and TANTO for adversarial detection without significantly compromising performance.

\begin{figure}[!h]
    \centering
    \includegraphics[width=0.4\textwidth]{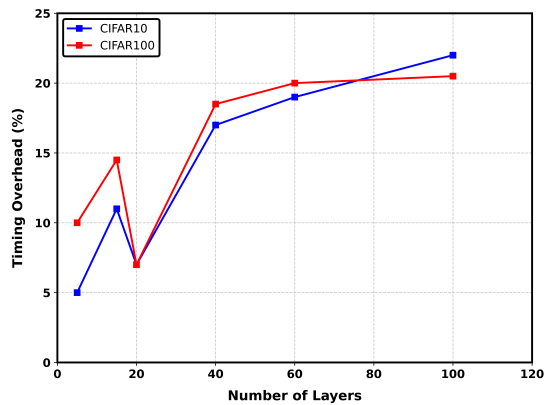}
    \caption {Timing overhead vs number of layers.}

    \label{fig:18}
\end{figure}




Figure \ref{fig:18} graphically represents the overhead results of the integrated APC and TANTO combination on different models and data sets. The figure \ref{fig:18} shows that the overhead varies depending on the model and dataset, aligning with the data presented in Table \ref{tab:my-table-11}. 
\section{\textbf{Conclusions}}
We have presented \textit{SAMURAI} -- a framework for protecting AI accelerators against major adversarial attack vectors by integrating performance counters (APC) and a runtime attack detection block called \textit{TANTO} that acts on the recorded performance counter traces. \textit{SAMURAI} provides enhanced security, efficiency, and integrity of AI operations at low overhead through runtime monitoring of internal events. As APC records the low-level parameter traces of AI operation, the small or insignificant perturbation, which forces the ML model to make incorrect predication is accurately detected by this technique. The on-device learning and inferencing in TANTO lead to minimal data transfer overhead, i.e., communication bandwidth while optimizing storage and processing resources, while providing real-time attack detection capability. The experimental results on various AI models using different datasets show that the proposed hardware solution has achieved approximately 97\% attack detection accuracy with modest overhead. Our future work will focus on the implementation of energy and area-efficient APCs for enhanced threat detection, extension to more complex AI models, including generative AI, and scalable cloud-based security solutions.


\bibliographystyle{IEEEtran}
\bibliography{RTL_conf}

\end{document}